\documentclass{aa}
\newcommand{\EQ}{\begin{equation}}
\newcommand{\EN}{\end{equation}}
\newcommand{\EQA}{\begin{eqnarray}}
\newcommand{\ENA}{\end{eqnarray}}
\newcommand{\eq}[1]{(\ref{#1})}
\newcommand{\Eq}[1]{Eq.~(\ref{#1})}
\newcommand{\Eqs}[2]{Eqs.~(\ref{#1}) and~(\ref{#2})}

\newcommand{\App}[1]{Appendix~\ref{#1}}
\newcommand{\Sec}[1]{Sect.~\ref{#1}}
\newcommand{\Fig}[1]{Fig.~\ref{#1}}

\newcommand{\Tab}[1]{Table~\ref{#1}}
\newcommand{\Figs}[2]{Figs~\ref{#1} and \ref{#2}}
\newcommand{\Tabs}[2]{Tables~\ref{#1} and \ref{#2}}
\newcommand{\bra}[1]{\langle #1\rangle}

\newcommand{\meanBB}{\overline{\mbox{\boldmath $B$}}{}}
\newcommand{\meanJJ}{\overline{\mbox{\boldmath $J$}}{}}

\newcommand{\uu}{\mbox{\boldmath $u$} {}}

\newcommand{\BB}{\mbox{\boldmath $B$} {}}

\newcommand{\AAA}{\mbox{\boldmath $A$} {}}

\newcommand{\JJ}{\mbox{\boldmath $J$} {}}

\newcommand{\ff}{\mbox{\boldmath $f$} {}}

\newcommand{\grav}{\mbox{\boldmath $g$} {}}
\newcommand{\nab}{\mbox{\boldmath $\nabla$} {}}

\newcommand{\ii}{{\rm i}}

\newcommand{\DD}{{\rm D} \, {}}
\newcommand{\dd}{{\rm d} {}}

\def\la{\mathrel{\mathchoice {\vcenter{\offinterlineskip\halign{\hfil
$\displaystyle##$\hfil\cr<\cr\sim\cr}}}
{\vcenter{\offinterlineskip\halign{\hfil$\textstyle##$\hfil\cr<\cr\sim\cr}}}
{\vcenter{\offinterlineskip\halign{\hfil$\scriptstyle##$\hfil\cr<\cr\sim\cr}}}
{\vcenter{\offinterlineskip\halign{\hfil$\scriptscriptstyle##$\hfil\cr<\cr\sim\cr}}}}}

\newcommand{\ea}{{\rm et al. }}

\def\half{{\textstyle{1\over2}}}

\def\onethird{{\textstyle{1\over3}}}

\def\quarter{{\textstyle{1\over4}}}
\newcommand{\yjgr}[3]{, #1, {JGR }{#2}, #3}
\newcommand{\yapj}[3]{, #1, {ApJ }{#2}, #3}

\newcommand{\yana}[3]{, #1, {A\&A }{#2}, #3}

\newcommand{\yjfm}[3]{, #1, {JFM }{#2}, #3}

\newcommand{\ypr}[3]{, #1, {Phys. Rev. } {#2}, #3}

\newcommand{\ybook}[3]{, #1, {#2} (#3)}

\newcommand{\pproc}[4]{, #1, in {#2}, ed. #3 (#4, in press)}
\newcommand{\ppp}[1]{, #1, {Phys. Plasmas, submitted}}

\newcommand{\smn}[1]{, #1, {MNRAS } (submitted)}

\input{epsf}
\begin{document}
\thesaurus{12(02.08.01; 02.13.1; 02.13.2; 02.20.1; 08.13.1; 11.13.2)}
\title{Large scale dynamos with helicity loss through boundaries}
\author{Axel Brandenburg\thanks{Also at: NORDITA, Blegdamsvej 17, DK-2100 Copenhagen \O, Denmark}
and Wolfgang Dobler}
\institute{
Department of Mathematics, University of Newcastle upon Tyne, NE1 7RU, UK}

\maketitle

\begin{abstract}
Dynamo action is investigated in simulations of locally isotropic and homogeneous
turbulence in a slab between open boundaries. It is found that a
``pseudo-vacuum'' boundary condition (where the field is vertical) leads to strong helicity
fluxes which significantly reduce the amplitude of the resulting large
scale field. On the other hand, if there is a conducting halo outside the
dynamo-active region the large scale field amplitude can reach larger
values, but the time scale after which this field is reached increases
linearly with the magnetic Reynolds number. In both cases most of the
helicity flux is found to occur on large scales. From the variety of models
considered we conclude that open boundaries tend to lower the saturation
field strength compared to the case with periodic boundaries. The rate
at which this lower saturation field strength is attained is roughly
independent of the strength of the turbulence and of the boundary conditions.
For dynamos with less helicity, however, significant field strengths
could be reached in a shorter time.
\end{abstract}

\section{Introduction}

Significant progress has been made in recent years in the understanding
of the generation of large scale magnetic fields in astrophysical bodies
by dynamo action. Some essential features of mean-field dynamo theory
(e.g., Moffatt 1978, Krause \& R\"adler 1980) are recovered qualitatively
and quantitatively in direct three-dimensional simulations, but the time
scale on which large scale fields are established becomes very long
as the magnetic Reynolds number increases (Brandenburg 2001a, hereafter
referred to as B2001). This result, which has been obtained using simulations
of helical turbulence in a periodic domain, is now understood to be a
consequence of the fact that the large scale magnetic fields generated
by such flows possess magnetic helicity which is conserved, except
for (microscopic) resistive effects. This behaviour
can also be reproduced with a mean-field model with resistively dominated
(`catastrophic') quenching of $\alpha$-effect and turbulent magnetic
diffusivity; see B2001 for details.
The implications of boundary conditions for `catastrophic'
$\alpha$-quenching were first pointed out by Blackman \& Field (2000a),
who showed that the results of earlier simulations by Cattaneo \& Hughes (1996)
could be understood as a consequence of periodic boundary conditions.

The problem is that in order to generate large scale fields with finite
magnetic helicity 
on time scales shorter than the resistive time,
one also needs to generate magnetic fields with the
opposite sign of magnetic helicity, so that
the net magnetic helicity stays close to its initial value.
In an astrophysical body the sign of the magnetic helicity is different
in the two hemispheres,
but preliminary models with helicity
flux through the equatorial plane suggest that this actually diminishes
the net large scale
magnetic field rather than enhancing it (Brandenburg 2001b). The other
possibility is that one can have a segregation in scales
(e.g.\ Seehafer 1996) such that large
scale magnetic fields with magnetic helicity of one sign are generated
within the domain of interest, whilst small scale magnetic fields with
opposite sign of magnetic helicity are expelled through the boundaries
(Blackman \& Field 2000b).
Whether or not this mechanism really works is unclear. So far there has only
been a phenomenological approach suggesting that the $\alpha$-effect
would be large if a flux of magnetic helicity through the boundaries was
possible (Kleeorin \ea 2000).

The aim of the present paper is to investigate the effect of open
boundaries on the dynamo.
A simple local boundary condition is the vertical field
condition (sometimes also referred to as pseudo-vacuum boundary
condition), that was used previously both in magnetoconvection
(Hurlburt \& Toomre 1988) and in accretion disc simulations showing
dynamo action (Brandenburg \ea 1995). This
boundary condition allows for a finite magnetic helicity flux
through the boundaries, even though the flux of magnetic energy (i.e.\
the Poynting flux) is still zero. We also consider cases with a conducting
halo outside the disc, as well as cases with density stratification,
so that magnetic buoyancy can contribute to turbulent flux transport out of
the dynamo-active region.

\section{The model}
\label{Smodel}

We solve the isothermal compressible MHD equations for the logarithmic
density $\ln\rho$, the velocity $\uu$, and the magnetic vector potential
$\AAA$,
\EQ
{\DD\ln\rho\over\DD t}=-\nab\cdot\uu,
\EN
\EQ
{\DD\uu\over\DD t}=-c_{\rm s}^2\nab\ln\rho+{\JJ\times\BB\over\rho}
+{\mu\over\rho}(\nabla^2\uu+\onethird\nab\nab\cdot\uu)+\ff,
\label{dudt}
\EN
\EQ
{\partial\AAA\over\partial t}=\uu\times\BB-\eta\mu_0\JJ,
\label{dAdt}
\EN
where ${\rm D}/{\rm D}t=\partial/\partial t+\uu\cdot\nab$ is the
advective derivative, $\BB=\nab\times\AAA$ the magnetic field,
$\JJ=\nab\times\BB/\mu_0$ the current density, and $\ff$ is the
random forcing function specified in B2001.
The induction equation (\ref{dAdt}) implies a specific gauge
for $\AAA$; in \Sec{Shelic} we discuss how to calculate gauge-independent
magnetic helicity and helicity fluxes. Instead of the dynamical viscosity
$\mu$ ($=\mbox{const}$) we will in the following refer to
$\nu\equiv\mu/\rho_0$, where $\rho_0$ is the mean density in the
domain ($\rho_0=\mbox{const}$ owing to mass conservation).
We use nondimensional units where $c_{\rm s}=k_1=\rho_0=\mu_0=1$.
Here, $c_{\rm s}=\mbox{const}$ is the sound speed, $k_1$ the smallest
wavenumber of the domain (so its size is $2\pi$),
and $\mu_0$ is the vacuum permeability.

We use sixth-order finite difference and a third-order time stepping
scheme. The number of meshpoints is usually $120^3$, except
for a few low Reynolds number runs.

All variables are assumed to be periodic in the $x$- and $y$-directions.
This implies that the vertical magnetic flux vanishes, because
$B_z=\partial_x A_y-\partial_y A_x$, so the integral of $B_z$ over $x$
and $y$ vanishes. Therefore the horizontal average of the vertical field
also vanishes. In the $z$-direction we assume
either also periodicity (if there is a halo), or we adopt a stress-free
boundary condition where the field is purely vertical (if there is
no halo). The latter states that $\partial_z u_x=\partial_z u_y=u_z=0$
together with $B_x=B_y=0$, and solenoidality then implies that $\partial_z
B_z=0$ on the vertical boundaries. In terms of $\AAA$, this boundary
condition is $\partial_z A_x=\partial_z A_y=A_z=0$.
In models with a halo the extent in the $z$-direction is $4\pi$ and
$\ff$ is multiplied by a mask function
\EQ
p(z)=\left\{
\begin{array}{ll}
\exp[-z^2/(H^2-z^2)]\quad&\mbox{for}\quad|z|<H,\\
0                        &\mbox{for}\quad|z|\ge H,
\end{array}
\right.
\EN
where $H=\pi$ is the semi-height of the disc. In the models with halo we
adopt periodic boundary conditions which has some conceptual advantages
in that all the magnetic energy and helicity that crosses the boundary at
$z=\pm H$ can be accounted for. We also consider models where, in addition
to the halo, we have
imposed gravitational acceleration, so $\ff\rightarrow\ff+\grav$, where
$\grav=-\nab\Phi$ and $\Phi=-g_0\cos(z/2)$ is a periodic gravitational potential
in $|z|<2\pi$. This produces a density enhancement near the midplane at $z=0$.
The models presented below have $g_0=0.5$, so the maximum density contrast
is $\Delta\ln\rho=2g_0/c_{\rm s}^2=1$.
These models allow for additional flux loss via magnetic buoyancy. We
first discuss the models without halo and uniform background density.

\section{Description of the various runs}

The parameters of the various runs are summarized in \Tab{T1}. All runs develop a large
scale magnetic field after some time. This large-scale field points in
the horizontal directions and varies along $z$. However, in some cases
there can be intermediate stages during which the large scale field is
oriented differently. The different orientations of the mean field are
best described by calculating the three different mean fields, whose
energies are denoted in the following by $K_x$, $K_y$, and $K_z$. The
corresponding mean fields vary only in the $x$, $y$, and $z$ directions,
respectively, and are averaged in the two perpendicular directions;
see also Eqs~(8)--(10) of B2001. The mean field that varies only in
the $z$-direction, and which is averaged over $x$ and $y$, is denoted
in the following by an overbar.

\begin{table}[t!]\caption{
Summary of the main properties of the runs with vertical field
boundary condition and with conducting halo. The models Buoy~1 and Buoy~2
are stratified and allow for magnetic buoyancy effects. The parameter
$q=\bra{\meanBB^2}/\bra{\BB^2}$ gives the fractional magnetic
energy in the mean field relative to the total magnetic field.
}\vspace{12pt}\centerline{\begin{tabular}{ccccccc}
      &  $\nu$ & $\eta$&$\eta_{\rm halo}$&$u_{\rm rms}$&$b_{\rm rms}$&  $q$  \\
\hline
Vert~1 &  0.01  &  0.01   &      --      &   0.100    &    0.093    &  0.76 \\
Vert~2 &  0.005 &  0.005  &      --      &   0.161    &    0.141    &  0.59 \\
Vert~3 &  0.002 &  0.002  &      --      &   0.197    &    0.178    &  0.39 \\
Vert~4 &  0.002 &  0.001  &      --      &   0.191    &    0.166    &  0.24 \\
Halo~1 &  0.002 &  0.0005 &     0.02     &   0.116    &    0.137    &  0.44 \\
Halo~2 &  0.002 &  0.001  &     0.02     &   0.118    &    0.156    &  0.60 \\
Buoy~1 &  0.002 &  0.001  &     0.02     &   0.124    &    0.200    &  0.61 \\
Buoy~2 &  0.002 &  0.001  &     0.001    &   0.147    &    0.242    &  0.71 \\
\label{T1}\end{tabular}}\end{table}

\subsection{Runs with vertical field boundary condition}

\begin{figure}[t!]\epsfbox{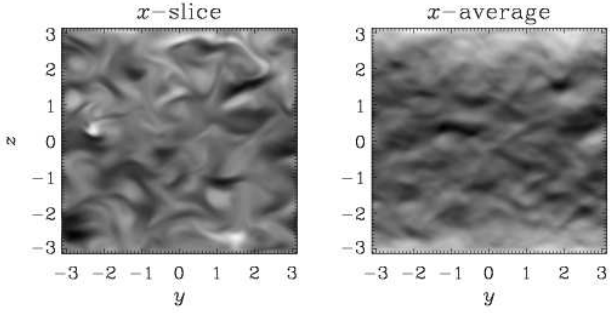}\caption[]{
Images of a cross-section and an $x$-average of $B_y$.
Vert~4, $t=1150$. Dark and light indicate shades indicate
negative and positive values, respectively.
Note the high noise level in the cross-section.
}\label{Fpslice_mhdpx}\end{figure}

In \Fig{Fpslice_mhdpx} we show the resulting field structure near the
end of run Vert~4 at $t=1150$. Unlike the case with periodic boundaries
where the presence of a large scale field was readily visible in
two-dimensional slices of individual field components (see Fig.~4 in
B2001), averaging over one of the two horizontal coordinate
directions is required before the mean field becomes visible.

\epsfxsize=8.2cm

\begin{figure}[t!]\epsfbox{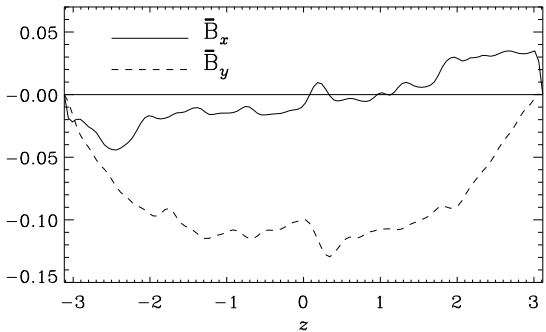}\caption[]{
Horizontal averages of $B_x$ and $B_y$.
Vert~4.
}\label{Fpslice_aver_mhdpx}\end{figure}

Averaging over both horizontal coordinate directions clearly yields the
profiles of the two non-vanishing field components, ${\overline B}_x$
and ${\overline B}_y$; see \Fig{Fpslice_aver_mhdpx}. Note that one of the two
components (here ${\overline B}_x$) is approximately antisymmetric about
the midplane, whilst the other (here ${\overline B}_y$) is approximately
symmetric about the midplane and of larger amplitude. There are also
some indications of boundary layer behaviour which was seen in highly
nonlinear $\alpha^2$-dynamos (Meinel \& Brandenburg 1990), with the
same boundary conditions as in the present simulations, i.e.\
$\overline{B}_x=\overline{B}_y=0$.

\epsfxsize=8.2cm

\begin{figure}[t!]\epsfbox{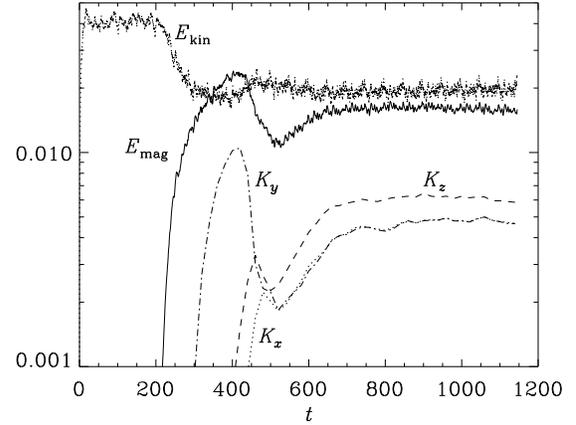}\caption[]{
Evolution of kinetic and magnetic energy together with the magnetic
energies contained in three differently averaged mean fields.
Vert~3.
}\label{FRR1t}\end{figure}

\begin{figure}[t!]\epsfbox{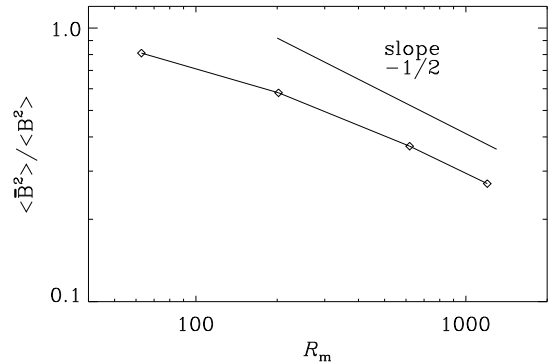}\caption[]{
Mean-field magnetic energy relative to the total magnetic energy,
$q=\bra{\meanBB^2}/\bra{\BB^2}$, for different values of the
magnetic Reynolds number, $R_{\rm m}=u_{\rm rms}L/\eta$, where
$L=2\pi$ is the scale of the simulation domain.
}\label{Fpq}\end{figure}

In \Fig{FRR1t} we plot the evolution of the magnetic energy within the
domain. The magnetic diffusion time is $(\eta k_1^2)^{-1}\approx500$.
For $t\la200$, magnetic energy is mainly in the small scales, for which
the helicity constraint is irrelevant, so the initial exponential growth
can occur on a dynamical time scale. At $t=400$, the magnetic energy
reaches a peak.
At that time most of the large scale field energy is actually
in $K_y$, which denotes the energy of the mean field defined by averaging
in the two directions perpendicular to $y$, so this field varies only in
the $y$-direction. This field is different from the anti\-ci\-pated field
$K_z$ that is obtained by averaging in the $x$ and $y$-directions.
This $K_z$ field also builds up and saturates around $t=700$, but it
reaches a saturation level that is significantly less than in the
case of periodic domains (B2001). Furthermore, as the magnetic Reynolds
number increases, the fractional energy of the mean field relative
to the total magnetic field decreases; see \Fig{Fpq}.

\epsfxsize=8.2cm

\epsfxsize=8.2cm

\subsection{Runs with a conducting halo}

We have carried out runs with low and high halo conductivities, with
and without imposed gravity. All these runs are similar with respect to
the overall field evolution in that they all produce a clearly visible
large scale field that grows slowly (on a resistive time scale), reaching
finally super-equipartition field strengths. These runs are therefore
similar to those of periodic domains, but rather different to those with
vertical field boundary conditions.

\begin{figure}[t!]\epsfbox{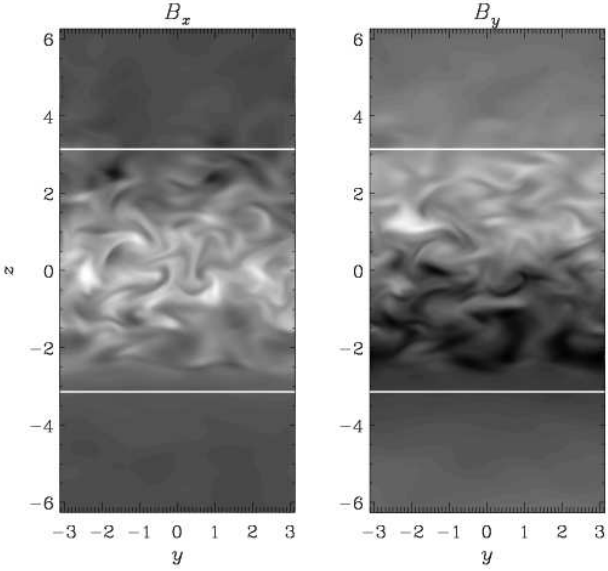}\caption[]{
Slices of the two horizontal field components for Buoy~2 with conducting
halo and vertical gravity. The boundaries between halo and disc plane are
indicated by white lines. $t=1400$. Dark and light shades
indicate negative and positive values, respectively.
}\label{Fpslice_mhdp}\end{figure}

\begin{figure}[t!]\epsfbox{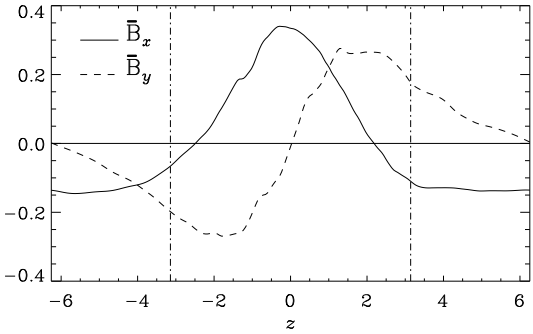}\caption[]{
Horizontal averages of $B_x$ and $B_y$ for Buoy~2. The boundary between halo and disc
plane is indicated by the vertical dash-dotted lines. $t=1400$. Note that at this
late time significant amounts of magnetic flux have been diffused into the halo.
}\label{Fpslice_aver_mhdp}\end{figure}

In \Fig{Fpslice_mhdp} we show a vertical slice of the resulting field structure
for a run with gravity ($g_0=0.5$) and a halo conductivity that is equal to the
conductivity in the turbulent zone (in $|z|<\pi$). Note that, in contrast to 
\Fig{Fpslice_mhdpx}, the large scale field can clearly be seen even
without averaging. The horizontally averaged field components
are shown in \Fig{Fpslice_aver_mhdp}.

\epsfxsize=8.2cm

\epsfxsize=8.2cm

\begin{figure}[t!]\epsfbox{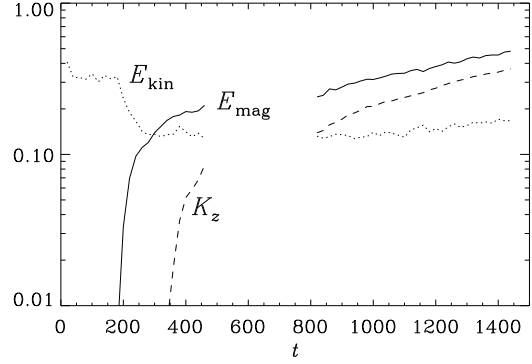}\caption[]{
Evolution of kinetic and magnetic energy together with the magnetic
energies of the mean field.
$\eta=10^{-3}$. Buoy~2. The second part of the run ($t\ge800$) was
obtained by restarting the simulation from Buoy~1.
}\label{FGGrav3t}\end{figure}

The evolution of kinetic and magnetic energies, as well as the magnetic
energy of the large scale field, are shown in \Fig{FGGrav3t} for Buoy~2. In addition
to a run starting with just a weak seed magnetic field we also show
how the energies evolve when restarting from a snapshot of Buoy~1 at $t=800$.

\epsfxsize=8.2cm

In the following we study the results obtained in view of the helicity
constraint, which was used extensively in B2001 in order to understand
the slow growth of the large scale field and its final saturation level.

\section{Helicity constraint and helicity fluxes}
\label{Shelic}

The resistively limited growth of large scale fields generated by helical
turbulence is primarily a consequence of helicity conservation. In the
present case, however, there is an additional surface term which
results from the flux of helicity passing through those boundaries, so
the equation of magnetic helicity conservation becomes
\EQ
{\dd H\over\dd t}=-2\eta\mu_0 C-Q,
\EN
where $H$ and $C$ are magnetic and current helicities, respectively, and
$Q$ is the surface-integrated helicity flux through the boundaries.

Owing to homogeneity of the system in the two horizontal directions we can
consider integral quantities normalized per unit surface, so the current
helicity (per unit surface) is therefore defined as
\EQ
C={1\over L_xL_y}\int\vec{J}\cdot\vec{B}\,\dd V\equiv
\int_{z_1}^{z_2}\overline{\vec{J}\cdot\vec{B}}\,\dd z,
\EN
where the overbar denotes $x$ and $y$ integration and division by the
corresponding surface area $L_xL_y$ ($=4\pi^2$), i.e.\ a horizontal average. The
magnetic helicity would be $\int\overline{\vec{A}\cdot\vec{B}}\,\dd z$,
but in this form it is not gauge-invariant and would change if a constant
or some gradient field were added to $\AAA$, which would leave $\BB$
unchanged\footnote{We recall that in a periodic domain,
$\int\vec{A}\cdot\vec{B}\,\dd V$ is automatically gauge invariant,
because then there are no surface terms and
$\int\vec{\nabla}\phi\cdot\vec{B}\,\dd V=
-\int\phi\vec{\nabla}\cdot\vec{B}\,\dd V=0$ for any $\phi$.}.
The gauge invariant helicity of Berger \& Field (1984) is
given by
\EQ
H=\int_{z_1}^{z_2}
{\overline{(\vec{A}+\vec{A}_0)\cdot(\vec{B}-\vec{B}_0)}}\,\dd z,
\label{gaugeinv}
\EN
where $\vec{B}_0=\vec{\nabla}\times\vec{A}_0$ is a potential field
that satisfies
\EQ
\nabla^2\vec{A}_0=\vec{0},\;\;
\vec{\nabla}\cdot\vec{A}_0=0,\;\;
A_{0z}=0\;\;\mbox{in}\;\; z_1<z<z_2,
\EN
and
\EQ
\vec{A}_0=\overline{\vec{A}}_0
+\vec{\nabla}_\perp\times(\psi\vec{\hat{z}}),\;\;
\vec{\nabla}_\perp^2\psi=-B_z\;\;\mbox{on}\;z=z_1,z_2;
\label{gaugeinv_bc}
\EN
here, $\vec{\nabla}_\perp=(\partial_x,\partial_y,0)$ is the
horizontal nabla operator. Like $\AAA$, we assume $\AAA_0$ and
$\psi$ to be periodic. This implies that the horizontal average of
$\vec{\nabla}_\perp\times(\psi\vec{\hat{z}})$ vanishes. This is the
reason why we had to include an explicit $\overline{\vec{A}}_0$ term
in \Eq{gaugeinv_bc} which is not needed by Berger \& Field (1984).
Since the gauge-invariant magnetic helicity of the
mean field plays an important role in the present context, we
give an explicit derivation of this component of the magnetic
helicity in \App{Sapp}.

\begin{figure}[t!]\epsfbox{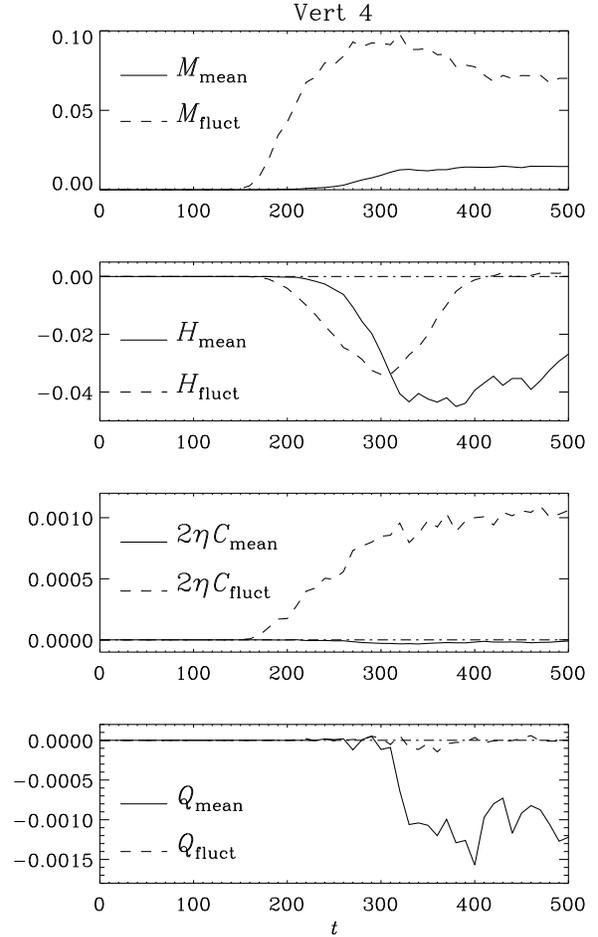}\caption[]{
Evolution of magnetic energy $M$, magnetic helicity $H$, current helicity
$2\eta C$ and magnetic helicity flux $Q$, for mean (solid lines) and
fluctuating (dashed lines) field components for Vert~4.
($\mu_0=1$ factors are omitted in the legends.)
}\label{FphelRR2}\end{figure}

For the following analysis it is convenient to split the helicity terms
into contributions from mean and fluctuating fields,
$H=H_{\rm mean}+H_{\rm fluct}$, with [see \Eq{ginv_Hmean}]
\EQ
H_{\rm mean}=\int_{z_1}^{z_2}
\overline{\vec{A}}\cdot\overline{\vec{B}}\,\dd z
+\hat{\vec{z}}\cdot(\overline{\vec{A}}_1\times\overline{\vec{A}}_2),
\label{Hmean}
\EN
where $\overline{\vec{A}}_1$ and $\overline{\vec{A}}_2$ are the values
of $\overline{\vec{A}}$ at $z=z_1$ and $z_2$, respectively, and
\EQ
H_{\rm fluct}=\int_{z_1}^{z_2}
{\overline{(\vec{a}+\vec{a}_0)\cdot(\vec{b}-\vec{b}_0)}}\,\dd z.
\label{Hfluct}
\EN
The latter expression is just \Eq{gaugeinv}, but rewritten for
the fluctuating fields, $\vec{a}=\vec{A}-\overline{\vec{A}}$, and
correspondingly for the other variables. [We recall that \Eq{Hmean} can
also be written in a form similar to \Eq{Hfluct}.] The components of
the helicity flux, $Q=Q_{\rm mean}+Q_{\rm fluct}$, are given by
(see \App{Sapp2})
\EQ
Q_{\rm mean}=-(\overline{\vec{E}}_1+\overline{\vec{E}}_2)\cdot
\int_{z_1}^{z_2}\overline{\vec{B}}\,\dd z
\EN
and (see Berger \& Field 1984)
\EQ
Q_{\rm fluct}=2\hat{\vec{z}}\cdot
(\overline{\vec{e}\times\vec{a}_0})\,\bigr|_{z_1}^{z_2},
\EN
where
\EQ
\vec{E}=\eta\mu_0\vec{J}-\vec{u}\times\vec{B}
\EN
is the electric field, and $\vec{E}=\overline{\vec{E}}+\vec{e}$
is its decomposition into mean and fluctuating components.
In \App{Sapp3} we give the values of magnetic energy, magnetic helicity,
integrated helicity flux, as well as the current helicity, for the
eigenfunction of the $\alpha^2$-dynamo with the boundary condition
$\overline{B}_x=\overline{B}_y=0$. Next we present in \Fig{FphelRR2}
the evolution of magnetic energy, magnetic and current helicities,
and integrated helicity flux, for model Vert 4.
Note that, in the steady
state, the helicity flux is balanced by the current helicity term, i.e.\
\EQ
2\eta\mu_0 C_{\rm fluct}\approx-Q_{\rm mean}.
\label{balance}
\EN
The small scale current helicity term, $2\eta\mu_0 C_{\rm fluct}$,
may therefore be regarded as the source of magnetic helicity. Somewhat
surprisingly, however, most of the helicity flux that is carried away
results from the mean field, whilst most of the current helicity comes
from the fluctuations. Similarly, $|H_{\rm mean}|\gg|H_{\rm fluct}|$. This
dominance is partly explained by the factor $(k_{\rm f}/k_1)^2=25$
by which the large scales would exceed the small scales in the magnetic
helicity relative to the current helicity [see B2001 after Eq.~(41)].

\epsfxsize=8.2cm

\begin{table}[t!]\caption{
Comparison of relevant quantities in the magnetic helicity equation
for different runs with vertical field boundary conditions.
}\vspace{12pt}\centerline{\begin{tabular}{lllccccc}
                              &   Vert~1 &  Vert~2 &  Vert~3 &  Vert~4 \\ 
\hline
mesh                          &   $30^3$ &  $60^3$ & $120^3$ & $120^3$ \\
$\eta$                        &   0.01   &  0.005  &  0.002  &  0.001  \\
$M_{\rm mean}$                &   0.021  &  0.037  &  0.038  &  0.021  \\
$-H_{\rm mean}$               &   0.051  &  0.069  &  0.084  &  0.037  \\
$2\eta C_{\rm fluct}$         &   0.0012 &  0.0023 &  0.0021 & 0.0012  \\
$-Q_{\rm mean}$               &   0.0007 &  0.0017 &  0.0017 & 0.0010  \\
$M_{\rm fluct}/M_{\rm mean}$  &    0.3   &   0.7   &   1.6   &   3.2   \\
$H_{\rm fluct}/H_{\rm mean}$  &  $-0.03$ & $-0.09$ & $-0.12$ & $-0.23$ \\
$C_{\rm fluct}/C_{\rm mean}$  &   $-4$   &  $-13$  &  $-26$  &  $-69$  \\
$Q_{\rm fluct}/Q_{\rm mean}$  &  $-0.000$& $-0.003$& $-0.005$& $-0.007$\\
$\varepsilon_H$               &  $-2.5$  &  $-1.9$ &  $-2.2$ &  $-1.8$ \\
$\varepsilon_C$               &  $-0.7$  &  $-0.5$ &  $-0.5$ &  $-0.4$ \\
$\varepsilon_Q$               & $-0.35$  & $-0.29$ & $-0.23$ & $-0.26$ \\
\label{T2}\end{tabular}}\end{table}

It is then not too surprising, although perhaps somewhat disappointing,
that almost all the magnetic helicity flux is in the large scales. 
We note, however, that the fractional contribution of small scale helicity
flux does increase somewhat with increasing magnetic Reynolds number; see \Tab{T2},
where quantities relevant for the magnetic helicity equation are
given for the different models with vertical field boundary condition.
The fact that the $Q$ and $C$ terms in \Eq{balance} do not balance exactly is
partly explained by the finite value of $dH/dt$ even towards the
end of the simulation.
A potentially important feature is the negative burst of small-scale magnetic
helicity around the time when the small scale field saturates
($t\approx300$ in \Figs{FphelRR2}{FphelDi5c}). In Halo~1 this is
also associated with a similar negative burst in $Q_{\rm fluct}$,
as well as a with a period of enhanced small scale and large
scale Poynting flux,
\EQ
P_{\rm fluct}=\hat{\vec{z}}\cdot(\overline{\vec{e}\times\vec{b}})
\,\bigr|_{z_1}^{z_2},\quad\mbox{and}\quad
P_{\rm mean}=\hat{\vec{z}}\cdot(\overline{\vec{E}}\times\overline{\vec{B}})
\,\bigr|_{z_1}^{z_2},
\EN
respectively. (We recall that in the case of the vertical field boundary
condition $P=0$ because $\BB$ is vertical.)

\begin{figure}[t!]\epsfbox{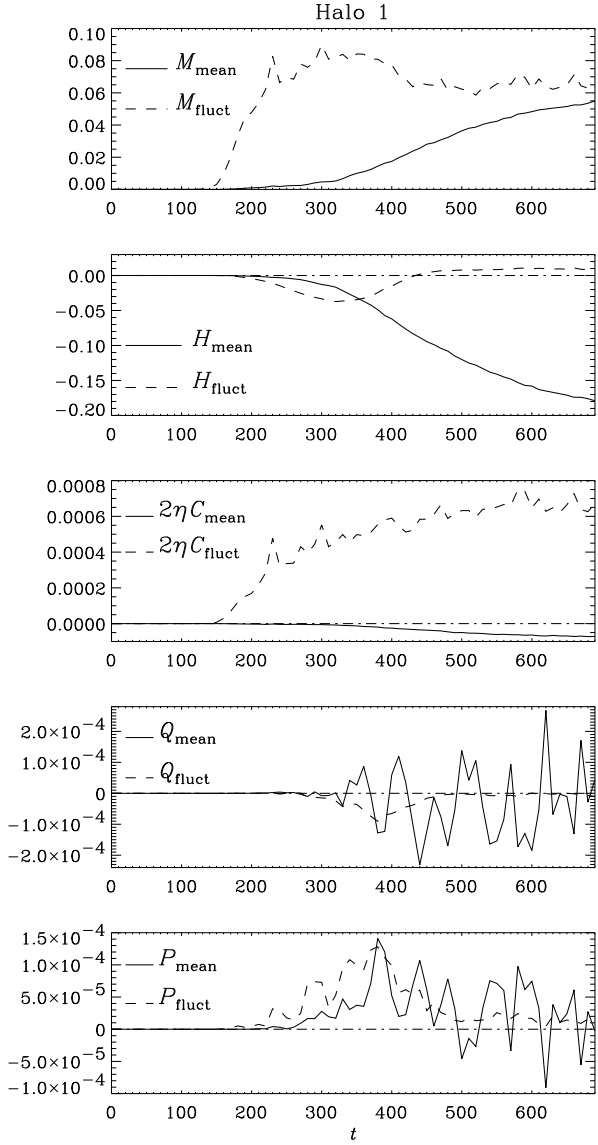}\caption[]{
Evolution of magnetic energy $M$, magnetic helicity $H$, current
helicity $2\eta C$, magnetic helicity flux $Q$, and magnetic
energy (or Poynting) flux $P$, for mean (solid lines) and fluctuating
(dashed lines) field components for Halo~1. All quantities are
evaluated between the boundaries $z_1=-\pi$ and $z_2=\pi$. After
$t\approx400\mbox{--}500$ the field in the halo becomes noticeable
and effects of the
periodic boundaries in the $z$-direction begin to affect the evolution
of the fluxes.
}\label{FphelDi5c}\end{figure}

Toward the end of the simulation with halo a significant mean field
has been accumulated in the halo (see \Fig{Fpslice_aver_mhdp}). This is
a somewhat unrealistic feature that results from the use of periodic
boundary conditions at the top of the halo. In reality the halo would
extend further and field would be lost into the exterior (i.e.\ interstellar
or intergalactic) medium. An additional artifact of the halo being
saturated with a large scale field is the fact that at late times the
Poynting flux is occasionally pointing inwards (\Fig{FphelDi5c}), which
is unlikely to occur in reality.

In \Tab{T2} we also give the helicities and corresponding fluxes of the
mean field in terms of the following nondimensional quantities:
\EQ
\varepsilon_H={k_1 H_{\rm mean}/\mu_0\over M_{\rm mean}},
\EN
\EQ
\varepsilon_C={C_{\rm mean}/k_1\over M_{\rm mean}},
\EN
\EQ
\varepsilon_Q={Q_{\rm mean}/\mu_0\over u_{\rm rms}M_{\rm mean}},
\EN
where $k_1=2\pi/L$ is the smallest wavenumber in the domain (and usually
the wavenumber of the large scale field); in our case $k_1=1$. We also have
$\mu_0=1$, but we keep this factor in some of the expressions for clarity.

\begin{table}[t!]\caption{
Comparison of relevant quantities in the magnetic helicity equation
for different runs with different implementations of a conducting
halo. Buoy~1 has a poorly conducting halo, whereas
Buoy~2 has a halo conductivity equal to the disc conductivity.
In Halo~1 the value of $\eta$ was so small that the full saturation
field strength has not yet been achieved by the end of the simulation. 
The asterisks mark those quantities that are likely to change if the run
was continued further. In all cases we use $120\times120\times240$
mesh points.
}\vspace{12pt}\centerline{\begin{tabular}{lllccccc}
                              &   Halo~1  &  Halo~2 &  Buoy~1 &  Buoy~2 \\ 
\hline
$\eta$                        &   0.0005 &  0.001  &  0.001  &  0.001  \\
$\eta_{\rm halo}$             &   0.0205 &  0.021  &  0.021  &  0.001  \\
$M_{\rm mean}$                & $0.052^*$&  0.092  &  0.15   &  0.26   \\
$-H_{\rm mean}$               & $0.170^*$&  0.300  &  0.53   &  0.94   \\
$2\eta C_{\rm fluct}$         &   0.0007 &  0.0012 &  0.0019 &  0.0021 \\
$-Q_{\rm mean}$               &   0.0000 &  0.0000 &  0.0007 &  0.0009 \\
$P_{\rm mean}$                &   0.0000 &  0.0000 &  0.0003 &  0.0003 \\
$M_{\rm fluct}/M_{\rm mean}$  &    1.3   &   0.7   &   0.6   &   0.4   \\
$H_{\rm fluct}/H_{\rm mean}$  &  $-0.06$ & $-0.04$ & $-0.03$ & $-0.02$ \\
$C_{\rm fluct}/C_{\rm mean}$  &   $-10$  &  $-4.8$ &  $-4.4$ &  $-3.1$ \\
$Q_{\rm fluct}/Q_{\rm mean}$  & $+0.27^*$& $-0.09$ & $+0.006$& $-0.034$\\
$\varepsilon_H$               &  $-3.3$  & $-3.3$  &  $-3.5$ &  $-3.6$ \\
$\varepsilon_C$               &  $-1.3$  & $-1.3$  &  $-1.4$ &  $-1.3$ \\
$\varepsilon_Q$               & $-0.002$ & $+0.005$& $-0.036$& $-0.022$\\
\label{T3}\end{tabular}}\end{table}

\epsfxsize=8.2cm

\Tab{T3} gives the same quantities for the model with a conducting
halo. Here the overall helicity flux is generally somewhat smaller
than in the case with vertical field boundary conditions. However,
the fractional contribution of small scale helicity flux is larger.
The magnetic helicity production, which is dominated by
$2\eta\mu_0 C_{\rm fluct}$, is still similar in the two cases.
In the following section we discuss why the losses of field with large
scale helicity significantly lower the saturation field strength.

\section{Interpretation}

\subsection{The helicity constraint}

In B2001 it was possible to understand the magnetic
field evolution as a consequence of the magnetic helicity evolution.
The current data show that most of the magnetic helicity and its corresponding
flux is in the mean field. Thus, helicity conservation is governed
approximately by
\EQ
{\dd\over\dd t}H_{\rm mean}\approx
-2\eta\mu_0(C_{\rm mean}+C_{\rm fluct})-Q_{\rm mean}.
\EN
Furthermore, in B2001 it was found that for strongly helical fields
both current and magnetic helicities are proportional to the large scale
magnetic energy. The same is true here: from \Tab{T2} we see that
the ratio $\varepsilon_H$ is around 2,
and that $\varepsilon_C$ is around 0.5--1,
We also note that
the helicity flux is a certain fraction of the magnetic energy, i.e.\
$\varepsilon_Q\approx\mbox{0.2--0.4}$.
Using the fact that all these ratios
are approximately constant for the different runs, we may write
\EQ
{\dd\over\dd t}M_{\rm mean}\approx
-2\eta_{\rm eff}k_1^2M_{\rm mean}
+2\eta k_1{C_{\rm fluct}\over\varepsilon_H},
\label{newdiff}
\EN
where we have defined an effective magnetic diffusion coefficient via
\EQ
2\eta_{\rm eff}=
(2\eta\varepsilon_C+\varepsilon_Qk_1^{-1}u_{\rm rms})/\varepsilon_H.
\EN
Equation \eq{newdiff} has the solution
\EQ
M_{\rm mean}\approx{\eta\over\eta_{\rm eff}}
{C_{\rm fluct}\over\varepsilon_H k_1}
\left[1-e^{-2\eta_{\rm eff}k_1^2(t-t_{\rm s})}\right]
\quad\mbox{for $t>t_{\rm s}$},
\label{helconstr}
\EN
where $t_{\rm s}=-\lambda\ln(B_{\rm ini}/B_{\rm eq})$ is the saturation
time of the total (mainly small scale) field, $\lambda$ is the
kinematic growth rate of the dynamo, and $B_{\rm ini}$ is the initial
field strength. Equation \eq{helconstr} is analogous to Eq.~(45) of B2001,
but since $\eta_{\rm eff}$ is much larger than $\eta$, the time scale
for reaching full saturation is reduced from $(\eta k_1^2)^{-1}$
to $(\eta_{\rm eff} k_1^2)^{-1}$ due to the presence of boundaries.
However, this result is obtained
at the price of having reduced the saturation field amplitude.
Therefore, all this flux term really does is cutting off the growth
of the field earlier than otherwise, so at any time the energy is
less than in the case of a periodic domain. Indeed, expanding the
exponential for short enough times we find that
\EQ
M_{\rm mean}\approx2\eta k_1{C_{\rm fluct}\over\varepsilon_H}(t-t_{\rm s})
\quad\mbox{for $t>t_{\rm s}$},
\label{helconstr2}
\EN
which shows that the initial growth time is still inversely proportional to
the microscopic magnetic diffusivity.

There is not much scope for increasing the rate of large scale field
production in \Eq{helconstr2}. The value of $C_{\rm fluct}$ does not
strongly depend on the details of the model (cf.\ \Tabs{T2}{T3}).
The only way out seems to be in lowering the value of $\varepsilon_{\rm H}$,
which corresponds to making the field less helical. This is what happened
when adding the effects of shear (Brandenburg \ea 2001).

If $\eta/\eta_{\rm eff}$ becomes very small the estimates given in
\Eqs{helconstr}{helconstr2} underestimate the value of $M_{\rm mean}$,
because there is always some mean field that results from imperfect
cancellations of the fluctuating field at the energy carrying wavenumber
which is here $k_{\rm f}$. The resulting energy of the mean field would
be of the order $(k_1/k_{\rm f})^3M_{\rm fluct}$. In the present case
we are however far away from this lower limit.

In the presence of a conducting halo the helicity flux is smaller
than in the case of the vertical field boundary condition; see \Tab{T3}
and \Fig{FphelDi5c} where we also show, in addition to magnetic
helicity flux and current helicity, the evolution of magnetic energy,
magnetic helicity, and the Poynting flux.
Note in particular that $Q_{\rm mean}$ now fluctuates about zero.
Consequently, there is no longer an
approximate balance between $2\eta C_{\rm fluct}$ and $Q_{\rm mean}$.
Instead, the $Q_{\rm mean}$ term becomes more important and would,
in the limit of a periodic domain ($Q\rightarrow0$), balance
$C_{\rm fluct}$, as was found in the case of periodic domains; see B2001.

\subsection{Comparison with a mean-field model}

In the case of a periodic domain it was possible to reproduce the slow
saturation of the magnetic energy with an $\alpha^2$-dynamo with
simultaneous $\alpha$ and $\eta_{\rm t}$-quenching of the form
\EQ
\alpha={\alpha_0\over1+\alpha_B\meanBB^2\!/B_{\rm eq}^2},\quad
\eta_{\rm t}={\eta_{\rm t0}\over1+\eta_B\meanBB^2\!/B_{\rm eq}^2},
\EN
where $\alpha_B=\eta_B$ was assumed, and both quantities were
proportional to the magnetic Reynolds number, as was suggested by
Vainshtein \& Cattaneo (1992). The reason this worked was because
of the presence of microscopic magnetic diffusion which was not
quenched, and which led to saturation such that [see Eqs~(55) and
(56) of B2001]
\EQ
\alpha_B{B_{\rm fin}^2\over B_{\rm eq}^2}\approx{\lambda\over\eta k_1^2}
\approx 0.01 R_{\rm m}
\quad\mbox{(for a periodic domain)},
\label{alpBper}
\EN
where $B_{\rm fin}$ is the final magnitude of the {\it mean} field,
$\lambda=\alpha k_1-\eta_{\rm T}k_1^2$ is the kinematic growth rate
of the dynamo, $\eta_{\rm T}=\eta+\eta_{\rm t}$ is the total magnetic
diffusivity, and $R_{\rm m}=u_{\rm rms}L/\eta$ is the magnetic Reynolds
number with respect to the scale of the domain.

\epsfxsize=8.2cm\begin{figure}[t!]\epsfbox{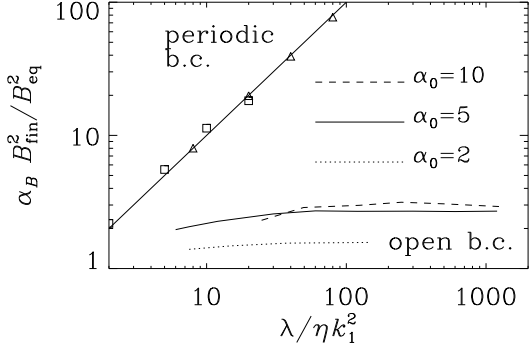}\caption[]{
Normalized magnetic energy in one-dimensional mean-field models with
periodic and open boundary conditions. The models with open boundaries
allow for helicity flux out of the domain and yield significantly smaller
field amplitudes when the magnetic Reynolds number, or $\lambda/\eta
k_1^2$, is large. In the periodic case the growth rate is
$\lambda=\alpha-\eta_{\rm T}$, whilst in the case with open boundaries
$\lambda=\quarter(\alpha^2-\eta_{\rm T}^2)/\eta_{\rm T}$.
For the periodic boundary condition, squares correspond to $\alpha_0=2$
and triangles to $\alpha_0=5$.
The data are obtained numerically using a time-stepping method.
}\label{Fpmfield}\end{figure}

In the present case of open boundaries there is an additional loss
term and so \Eq{alpBper} no longer holds. A numerical integration of
the nonlinear $\alpha^2$-dynamo equations with boundary conditions
$\overline{B}_x=\overline{B}_y=0$ (Meinel \& Brandenburg 1990) shows
that the term $\alpha_B B_{\rm fin}^2/B_{\rm eq}^2$ saturates at a value
of around 3 and is only weakly dependent on the value of $\alpha_0$;
see \Fig{Fpmfield}, where $\eta+\eta_{\rm t0}=1$. In the simulations,
on the other hand, the final large-scale field strength decreases with
$R_{\rm m}$ like $R_{\rm m}^{-1/2}$, even though the total (mainly small
scale) magnetic field strength is always close to the equipartition
field strength; see \Fig{Fpq} and \Tab{T1}. From this we can conclude
that $\alpha_B$
increases only like $R_{\rm m}^{1/2}$. This is to be contrasted with
the case of periodic domains where $B_{\rm fin}/B_{\rm eq}$ was
independent of $R_{\rm m}$ and therefore
$\alpha_B\propto R_{\rm m}$. It is
important to realize, however, that the scaling of $\alpha_B$ with
$R_{\rm m}$ is secondary to having strong large-scale dynamo action.
What really matters is the saturation level and the time scale on which
full saturation is reached. These properties are controlled primarily
by the magnetic helicity equation which, in turn, is sensitive to
boundary conditions, anisotropies, and the presence of shear, for
example (e.g.\ Brandenburg \ea 2001).

\section{Conclusions}

The results presented in this paper have shown that open boundary conditions, and
boundaries different from periodic ones, can have profound effects
on the large scale field generation in helical flows. Firstly,
unlike the case of periodic domains, a strictly force-free large scale
magnetic field is no longer possible. Secondly, the loss of magnetic
helicity through the boundaries severely limits the large scale field amplitude
attainable in simulations. This is quite different from the
case of periodic domains, where the large scale field energy can exceed the
kinetic energy, provided there is sufficient scale separation between
the energy carrying scale and the scale of the system
(see Sects~3.5 and 3.6 of B2001). For an open simulation volume
with a vertical field boundary condition, however, the large scale
field amplitude can be severely limited in a magnetic Reynolds number
($R_{\rm m}$) dependent fashion. Again, this result can be modelled
with a mean-field $\alpha^2$-dynamo, where the $\alpha$-effect and
turbulent magnetic diffusivity are quenched. However, the $R_{\rm
m}$-dependence of these quenching expressions is here less extreme
than in the case of periodic domains where the field can be
force-free (B2001). This suggests that a
dependence of $\alpha$ and $\eta_{\rm t}$ on $\meanBB^2$ alone is
too simplistic, and that there might be an additional dependence on
$\meanJJ\times\meanBB$ or $\meanJJ\cdot\meanBB$, for example,
that would react specifically on
non force-free magnetic fields.

Our result that the quenching of $\alpha$ and $\eta_{\rm t}$ can be
sensitive to external factors is not entirely unfamiliar.
For example in the presence of large scale shear (corresponding to
differential rotation) the quenching of $\alpha$ and $\eta_{\rm t}$
was already found to be significantly reduced; see Brandenburg \ea (2001).
A significant reduction of the quenching may explain why
the cycle period in oscillatory dynamos does not need to scale linearly
with the resistive time scale (which would be too long). Nevertheless,
the {\it growth time} of a large-scale dynamo with helicity
does seem to scale linearly with the
resistive time scale in all cases investigated so far.

In the simulations presented here the effects of open boundaries and
magnetic buoyancy were too weak to produce any sizeable small scale
magnetic helicity fluxes. However, this leaves unanswered the question
whether small scale magnetic helicity fluxes are {\it in principle} capable
of reducing the long time scales that result from magnetic helicity
conservation. In order to clarify this one may need to resort to models
where small scale flux losses are somehow artificially enhanced.

Finally, it should be mentioned that we have here neglected the effects
of an equator where the sign of the helicity would change. In the present
framework this can be modelled by modulating the sign of the helicity
of the forcing function in the $z$-direction. Preliminary results have
been reported in Brandenburg (2001b), but the results are quite similar
to those in the case of open boundary conditions. Indeed an estimate
similar to \Eq{helconstr} has been obtained for the resulting magnetic
energy of the mean field.

It is now important to move on to more realistic global
simulations which would {\it automatically} allow for helicity exchange across
the equator and produce flux losses through winds and eruptions into the
corona. It is indeed quite possible that significant helicity fluxes
are only produced by the combined action of stratification and shear
(Berger \& Ruzmaikin 2000, Vishniac \& Cho 2000), which again would
automatically be addressed by more realistic global simulations. Even
if the magnetic helicity flux continues to be dominated by large scale
contributions, it could be argued that such an effect could alleviate
the helicity constraint, because saturation (albeit at a lower value)
does occur earlier. It would seem necessary, however, to compensate those
additional losses by more powerful dynamo action. Again, the effects of
shear may prove important, because it provides additional toroidal field
that is non-helical and hence not subject to any constraint.

\begin{acknowledgements}
We thank Eric Blackman and Anvar Shukurov for critical comments and
suggestions that have improved the presentation of the results.
This work was partially supported by PPARC (Grant PPA/G/S/1997/00284)
and the Leverhulme Trust (Grant F/125/AL). Use of the PPARC supported
supercomputers in St Andrews and Leicester is acknowledged.
\end{acknowledgements}

\appendix
\section{Gauge-independent magnetic helicity of the mean field}
\label{Sapp}

In a periodic domain the  magnetic helicity is gauge invariant,
so $\overline{\vec{A}}\cdot\overline{\vec{B}}$ (in $z_1\leq z<z_2$)
can be made gauge invariant by linearly extrapolating
$\overline{\vec{A}}$ from $z_2$
to $z_3$, where $\overline{\vec{A}}_3=\overline{\vec{A}}_1$ is assumed,
and $\overline{\vec{A}}_i$ are the values of $\overline{\vec{A}}$ at
$z=z_i$, where $i=1$, 2, or 3. Linear extrapolation is used because
this corresponds to the simplest possible, current-free field. Thus,
we can write for the magnetic helicity of the mean field
\EQ
H_{\rm mean}=\int_{z_1}^{z_2}
\overline{\vec{A}}\cdot\overline{\vec{B}}\,\dd z
+\int_{z_2}^{z_3}\tilde{\vec{A}}\cdot\tilde{\vec{B}}\,\dd z,
\EN
where
\EQ
\tilde{\vec{A}}=\overline{\vec{A}}_2
+{\overline{\vec{A}}_3-\overline{\vec{A}}_2\over z_3-z_2}\,(z-z_2)
\EN
is the linear extension of $\overline{\vec{A}}$ from $z_2$ to $z_3$, and
\EQ
\tilde{\vec{B}}=\vec{\nabla}\times\tilde{\vec{A}}=\hat{\vec{z}}\times
{\overline{\vec{A}}_3-\overline{\vec{A}}_2\over z_3-z_2},
\EN
so
\EQ
\tilde{\vec{A}}\cdot\tilde{\vec{B}}=\overline{\vec{A}}_2\cdot
\left[\hat{\vec{z}}\times
{\overline{\vec{A}}_3-\overline{\vec{A}}_2\over z_3-z_2}\right]
=\hat{\vec{z}}\cdot
{\overline{\vec{A}}_3\times\overline{\vec{A}}_2\over z_3-z_2}.
\EN
Since $\overline{\vec{A}}_3=\overline{\vec{A}}_1$ we have
\EQ
\int_{z_2}^{z_3}\tilde{\vec{A}}\cdot\tilde{\vec{B}}\,\dd z
=\hat{\vec{z}}\cdot(\overline{\vec{A}}_1\times\overline{\vec{A}}_2),
\label{ginv_Hmean}
\EN
which is independent of the choice of the value of $z_3$. In particular,
we may choose $z_3=z_1$, in which case this definition is identical to
\Eq{gaugeinv} with \Eq{gaugeinv_bc} and $\overline{\vec{A}}_0=\tilde{\vec{A}}$.

\section{Helicity flux of the mean field}
\label{Sapp2}

The evolution of the gauge-invariant magnetic helicity of the mean
field is given by
\EQA
{\dd\over\dd t}H_{\rm mean}=
\int_{z_1}^{z_2}(\dot{\overline{\vec{A}}}\cdot\overline{\vec{B}}
+\overline{\vec{A}}\cdot\dot{\overline{\vec{B}}})\,\dd z
\nonumber\\
+\hat{\vec{z}}\cdot
(\dot{\overline{\vec{A}}}_1\times\overline{\vec{A}}_2
+\overline{\vec{A}}_1\times\dot{\overline{\vec{A}}}_2),
\ENA
where dots denote partial time derivatives,
but, because $\dot{\overline{\vec{A}}}=-\overline{\vec{E}}$ in our gauge
[cf.\ \Eq{dAdt}], we have
\EQA
{\dd\over\dd t}H_{\rm mean}=
-2\int_{z_1}^{z_2}\overline{\vec{E}}\cdot\overline{\vec{B}}\,\dd z
-\left.\hat{\vec{z}}\cdot(\overline{\vec{E}}\times\overline{\vec{A}})
\right|_{z_1}^{z_2}\nonumber\\
-\hat{\vec{z}}\cdot
(\overline{\vec{E}}_1\times\overline{\vec{A}}_2
+\overline{\vec{A}}_1\times\overline{\vec{E}}_2).
\ENA
The last two terms reduce to
\EQ
-\hat{\vec{z}}\cdot\left[
(\overline{\vec{E}}_1+\overline{\vec{E}}_2)\times
(\overline{\vec{A}}_2-\overline{\vec{A}}_1)\right],
\EN
which can also be written as
$(\overline{\vec{E}}_1+\overline{\vec{E}}_2)\cdot
\int_{z_1}^{z_2}\overline{\vec{B}}\,\dd z$.

\section{Helicity flux for the eigenfunction of the
associated $\alpha^2$-dynamo}
\label{Sapp3}

The kinematic $\alpha^2$-dynamo equation with
$\meanBB=\meanBB(z,t)=({\overline B}_x,{\overline B}_y,0)$
can be written in the form
\EQ
\partial_t{\cal B}=\ii\alpha\partial_z{\cal B}
+\eta_{\rm T}\partial_z^2{\cal B},
\EN
where we have adopted complex notation,
${\cal B}={\overline B}_x+\ii{\overline B}_y$, and
$\alpha$ and $\eta_{\rm T}$ are assumed constant.
The eigenfunction that satisfies ${\cal B}=0$ at $z=\pm\pi$ is
(e.g., Meinel \& Brandenburg 1990)
\EQ
{\cal B}(z,t)={\cal B}_0\cos{z\over2}\,
\exp\left(\lambda t-{\alpha\over\eta_{\rm T}}\,{\ii z\over2}\right),
\EN
where ${\cal B}_0$ is a complex constant and
$\lambda=\quarter(\alpha^2-\eta_{\rm T}^2)/\eta_{\rm T}$ is the growth
rate. We are interested in the marginally excited mode, $\lambda=0$,
and so we assume for simplicity ${\cal B}_0=\alpha=\eta_{\rm T}=1$, so
we have
\EQ
{\cal B}(z,t)=e^{-\ii z/2}\cos(z/2).
\EN
The complex current density, ${\cal J}=\ii\partial_z{\cal B}$,
is given by
\EQ
{\cal J}(z,t)={\textstyle{1\over2}}\exp(-\ii z),
\EN
and the complex vector potential is, apart from some arbitrary constant,
\EQ
{\cal A}(z,t)={\textstyle{1\over2}}\exp(-\ii z)
-{\textstyle{1\over2}}\ii z,
\EN
which satisfies $\partial_z^2{\cal A}=-{\cal J}$ and
${\cal B}=\ii\partial_z{\cal A}$. For this field the magnetic energy
$M_{\rm mean}$, the magnetic and current helicities $H_{\rm mean}$ and
$C_{\rm mean}$, as well as the
integrated magnetic helicity flux $Q_{\rm mean}$, are given by
\EQ
M_{\rm mean}={\pi\over2}=C_{\rm mean},\quad
H_{\rm mean}={3\pi\over2},\quad
Q_{\rm mean}=\pi.
\EN
The Lorentz
force is $\overline{\vec{J}}\times\overline{\vec{B}}=(0,0,\quarter\sin z)$,
and the current helicity density is 
$\overline{\vec{J}}\cdot\overline{\vec{B}}=\half\cos^2(z/2)$, so
\EQ
{\bra{(\overline{\vec{J}}\times\overline{\vec{B}})^2}\over
\bra{\overline{\vec{J}}^2}\bra{\overline{\vec{B}}^2}}={1\over4},\quad
{\bra{(\overline{\vec{J}}\cdot\overline{\vec{B}})^2}\over
\bra{\overline{\vec{J}}^2}\bra{\overline{\vec{B}}^2}}={3\over4},
\EN
so the field is neither fully force-free nor fully helical, but something
in between. (For the mean fields of B2001 those values were 0 and 1,
respectively.)

\end{document}